\documentclass[%
  aps,
  prapplied,
  twocolumn,
  superscriptaddress,
  longbibliography,
  floatfix,
  nofootinbib,
]{revtex4-2}

\usepackage{graphicx}
\usepackage{amsmath,amssymb}
\usepackage{bm}
\usepackage{booktabs}
\usepackage{braket}
\usepackage{capt-of}
\usepackage{xcolor}
\usepackage{tikz}
\usetikzlibrary{
  arrows.meta,
  positioning,
  fit,
  calc,
  decorations.pathreplacing,
  shapes.geometric,
  backgrounds,
}
\usepackage[colorlinks=true,linkcolor=blue,citecolor=blue,urlcolor=blue]{hyperref}

\makeatletter
\renewcommand\subsection{%
  \@startsection{subsection}{2}{\z@}%
    {1.5ex \@plus 0.4ex \@minus 0.2ex}%
    {-1sp}%
    {\normalfont\normalsize\itshape}%
}
\let\red@subsection\subsection
\renewcommand{\subsection}[1]{\red@subsection[#1]{#1.---}}
\makeatother

\newcommand{\RO}{\mathrm{RO}}

\newcommand{\affSQIL}{%
Superconductor Quantum Information Laboratory,
Institute of Physics,
{\'E}cole polytechnique f{\'e}d{\'e}rale de Lausanne (EPFL),
CH-1015 Lausanne, Switzerland}
\newcommand{\affAQUA}{%
Advanced Quantum Architecture Laboratory,
Institute of Electrical and Microengineering,
{\'E}cole polytechnique f{\'e}d{\'e}rale de Lausanne (EPFL),
CH-2000 Neuch{\^a}tel, Switzerland}

\begin{document}

\title{Syndrome-Resolved Decoding of Qubit Readout Errors}

\author{Han Zheng}
\thanks{These authors contributed equally to this work.}
\affiliation{\affSQIL}
\author{Vladimir Pe\v{s}i\'{c}}
\thanks{These authors contributed equally to this work.}
\affiliation{\affAQUA}
\author{Nikolay Gusarov}
\affiliation{\affSQIL}
\author{Andrea Lizzit}
\affiliation{\affSQIL}
\author{Taketo Imaizumi}
\affiliation{\affSQIL}
\author{Chulwoo Ahn}
\affiliation{\affSQIL}
\author{Marco Nicolas Mueller}
\affiliation{\affSQIL}
\author{Raymond Mencia}
\affiliation{\affSQIL}
\author{Edoardo Charbon}
\affiliation{\affAQUA}
\author{Vladimir E. Manucharyan}
\email{vladimir.manucharyan@epfl.ch}
\affiliation{\affSQIL}

\begin{abstract}
Readout errors in superconducting qubits include incorrect state assignment, bit flips within the computational subspace, and leakage to non-computational states. These errors require different mitigation strategies, yet conventional measurement fidelity and repeatability metrics do not resolve their individual contributions. Previous repeated-readout studies have shown that temporal correlations reveal additional information about readout errors. Here we introduce the Readout Error Decoder (RED), which converts repeated thresholded binary records into a channel-resolved error budget. Controlled error injections validate the channel-specific response of the decoder. We demonstrate RED on both fluxonium and transmon qubits and across distinct leakage-lifetime regimes. RED identifies the error channels limiting readout performance without requiring dedicated discrimination of non-computational states, providing a basis for targeted readout optimization across different qubit modalities.
\end{abstract}

\maketitle

\section{Introduction}
\label{sec:introduction}

An ideal qubit readout should correctly discriminate the qubit state without changing it~\cite{blais2021circuit, Lupascu2007QND}. This quantum-nondemolition (QND) property is essential for quantum error correction~\cite{google2021qec,krinner2022,zhao2022}, real-time feedback control~\cite{vijay2012feedback,campagne2013feedback,riste2013feedback}, and measurement-based quantum computation~\cite{raussendorf2001,briegel2009}. Conventional benchmarks quantify state discrimination through the signal-to-noise ratio (SNR) and measurement fidelity~\cite{walter2017,jeffrey2014,riste2012,swiadek2024,spring2026lowleakage}, together with repeatability as a practical QNDness benchmark~\cite{sunada2022,rilb,spring2025}. These scalar metrics are valuable measures of overall performance, but similar values can arise from distinct error mechanisms with different consequences and mitigation strategies. Identifying the dominant error channel is therefore important both for interpreting the readout error and for determining how the readout should be improved.
Recent repeated-readout studies have shown that correlations across successive readouts can reveal error information beyond conventional fidelity and repeatability metrics~\cite{rilb,high_freq_readout}. These results establish repeated readout as a promising route toward channel-resolved readout characterization.

A readout error in a superconducting qubit can arise through several mechanisms. An assignment error occurs when the qubit state is unchanged but the measured outcome is assigned incorrectly, for example because the $\ket{g}$ and $\ket{e}$ readout distributions overlap~\cite{walter2017,jeffrey2014}. A bit-flip error occurs when the qubit changes state within the computational subspace during the readout pulse~\cite{boissonneault2009,slichter2012,thorbeck2024}. Leakage occurs when population leaves the computational subspace~\cite{varbanov2020,mcewen2021,bista2026,nesterov2024}. Population can leak into different non-computational states, each with distinct readout characteristics and relaxation dynamics~\cite{sank2016mist,khezri2023,shillito2022,dumas2024ionization,cohen2023,wang2026probing}. We classify readout-induced leakage by its lifetime: long-lived leakage persists across multiple readouts~\cite{rilb}, whereas short-lived leakage can decay within a single readout duration but still disrupt an operation applied immediately after the readout. Among these error channels, leakage is particularly difficult to characterize. It may not be known a priori which leakage state is populated, where that state appears in the IQ plane, or how long the leaked population persists.

A common approach is to characterize leakage by resolving non-computational states through multi-state readout~\cite{khezri2023,high_freq_readout,wang2026probing,Wang2025HighEJEC}. This requires direct separation of the relevant higher-state responses in the IQ plane, which in practice depends on favorable state-dependent readout contrast and high SNR. Hazra \textit{et al.} introduced randomized interleaved readout benchmarking (RILB), which extracts leakage from the decay of repeated-readout correlations without resolving the leaked state directly~\cite{rilb}. However, RILB requires fast readout relative to the leakage lifetime, so that leaked population persists over successive readouts. Kurilovich \textit{et al.} combined repeated binary readout with a separate leakage check to resolve the different readout-error contributions~\cite{high_freq_readout}. These approaches therefore rely either on direct discrimination of higher states or on readout that is fast compared with the leakage dynamics. Neither condition is guaranteed in a general device. This leaves an open question: can readout error be decomposed into its underlying channels from repeated binary records across different devices and leakage regimes?

This inference problem has a close analogue in fault-tolerant quantum error correction, where repeated stabilizer measurements generate syndrome histories that are decoded to infer underlying errors~\cite{dennis2002,fowler2012}. Inspired by this logic, we introduce the Readout Error Decoder (RED), which converts repeated binary single-shot records into channel-specific error rates. Using known inter-readout gates, RED compares the measured bit string with the expected evolution and classifies the resulting disagreement patterns as syndromes. Long-lived leakage produces persistent correlation-breaking patterns, whereas short-lived leakage is probed with an inter-readout gate sequence that makes transient leakage visible in the binary record. RED therefore does not require leakage to persist over many readouts or the leakage states to be directly resolved in the IQ plane. The same framework also separates assignment and bit-flip errors according to their characteristic patterns in the repeated-readout record.

We validate RED using controlled bit-flip and leakage injections, which quantitatively test the extracted channel-specific error rates. We further demonstrate RED in both short- and long-lived leakage regimes using fluxonium and transmon devices, showing that the same decoding principle applies across different leakage timescales and qubit platforms. The resulting channel-resolved error budget identifies the limiting readout error channel and can guide targeted readout optimization~\cite{bengtsson2024,high_freq_readout}. Although demonstrated here on superconducting qubits, the underlying approach requires only repeated binary single-shot readout, known operations between readouts, and error processes that produce characteristic patterns across successive readouts. This suggests that RED could be adapted to other quantum platforms with repeated state detection and leakage outside a chosen computational subspace, including semiconductor spin qubits~\cite{xue2020prx,yoneda2020qnd}, trapped ions~\cite{hume2007qnd,negnevitsky2018}, and neutral-atom arrays~\cite{graham2023mcm,norcia2023mcm,zeng2026fndr}.

\section{Results}
\label{sec:results}

\subsection{RED protocol}
\label{sec:results:taxonomy}
\begin{figure*}
  \includegraphics[width=\textwidth]{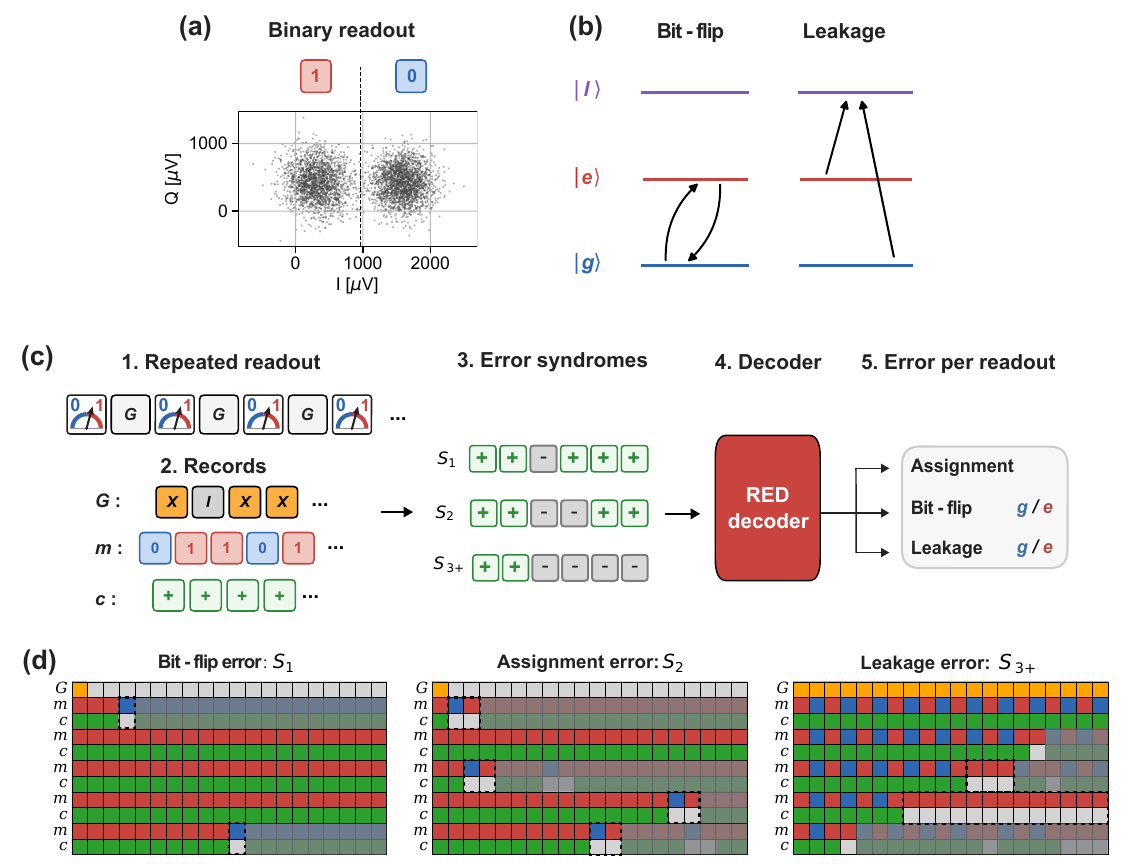}
  \caption{\textbf{Overview of the RED protocol.}
  \textbf{(a)}~Binary readout: each single-shot IQ point is thresholded into a bit $m_i\in\{0,1\}$.
  \textbf{(b)}~The two ways the qubit state can change during the readout: a flip inside the computational subspace, $\ket{g}\leftrightarrow\ket{e}$, and leakage to the non-computational state $\ket{l}$.
  \textbf{(c)}~The RED sequence and its analysis.
  Readouts are interleaved with a fixed gate $G\in\{I,X,X\text{-}\tau\text{-}X\}$, which gives the gate record $G$, the measured record $m$, and the correlation record $c$.
  Broken correlations are grouped by how many occur in a row into the syndromes $S_1$, $S_2$, and $S_{3+}$, and the decoder turns these counts into channel estimators for assignment, bit-flip, and leakage, resolved by the $\ket{g}$ and $\ket{e}$ branch. The quantitative channel extraction uses the decoder dictionary and experiment set of Tables~\ref{tab:decoder} and~\ref{tab:red_experiments}.
  \textbf{(d)}~Measured examples containing each syndrome (left to right): a bit-flip error ($S_1$), an assignment error ($S_2$), and leakage error ($S_{3+}$). Corresponding repeated-readout IQ traces examples are shown in Supplementary Fig.~\ref{fig:supp:qnd_traces}.
  Rows alternate between the gate $G$, the measured bits $m$, and the correlations $c$ (green: correlated, white: broken); the dashed boxes mark the first error in a row, and the shots after it are faded because they are discarded.  The $S_1$ and $S_2$ examples are from a fluxonium, and the $S_{3+}$ example is from a transmon.
  }
  \label{fig:error_taxonomy}
\end{figure*}

Figure~\ref{fig:error_taxonomy} provides an overview of RED, from binary single-shot readout to syndrome classification and error-channel extraction.
Conventional single-shot readout assigns each integrated IQ response to $\ket{g}$ or $\ket{e}$, as shown in Fig.\ref{fig:error_taxonomy}(a).
Its discrimination performance is characterized by the measurement fidelity,
$F_{\mathrm{m}}=\frac{1}{2}[P(0\mid g)+P(1\mid e)]$, where $P(0\mid g)$ and $P(1\mid e)$ are the probabilities of correctly assigning the prepared $\ket{g}$ and $\ket{e}$ states, respectively~\cite{walter2017,blais2021circuit}.
For repeated measurements, readout stability is commonly characterized by the branch-resolved repeatabilities
$R_g=P(0_2\mid 0_1,g)$ and $R_e=P(1_2\mid 1_1,e)$, or their average $R=(R_g+R_e)/2$, as a practical benchmark of QNDness~\cite{sunada2022,rilb}.

These two benchmarks quantify state discrimination and outcome stability, respectively, but do not identify the physical processes that limit them.
In addition to incorrect state assignment, readout can induce bit flips within the computational subspace, $\ket{g}\leftrightarrow\ket{e}$, or leakage to states outside it, as illustrated in Fig.~\ref{fig:error_taxonomy}(b).
Because assignment errors, bit flips, and leakage can produce the same binary outcome in a single readout, RED uses temporal correlations in the repeated binary record, together with the known inter-readout gates, to distinguish and quantify these error channels.

Figure~\ref{fig:error_taxonomy}(c) illustrates the RED workflow; a detailed description is provided in Supplementary Fig.~\ref{fig:supp:red_workflow}.
RED interleaves repeated readouts with a fixed operation
$G\in\{I,X,X\text{-}\tau\text{-}X\}$. 
Each gate is applied only after the readout resonator has rung down, waiting at least $10/\kappa$ before the next gate, where $\kappa$ is the resonator decay rate.
The prepared state and gate sequence define an expected logical record $l_i$, while thresholding each readout gives the measured record $m_i$.
The initial bit is set by the prepared state, $\ket{g}\!\to\!0$ or $\ket{e}\!\to\!1$.
Subsequent bits remain unchanged under $G=I$ or an ideal $X$-$\tau$-$X$ sequence and flip under $G=X$. RED compares changes in the expected and measured records through the correlation record
\begin{equation}
  C_i = 1 - 2\left[(l_{i-1}\oplus l_i)\oplus(m_{i-1}\oplus m_i)\right].
  \label{eq:correlation}
\end{equation}
Here $C_i=+1$ indicates agreement with the expected evolution, whereas $C_i=-1$ marks a broken correlation.
Consecutive broken correlations are grouped by run length into $S_1$, $S_2$, and $S_{3+}$, corresponding to one, two, and three or more broken correlations, respectively.

The run length reflects how long an error affects the binary record.
A bit flip changes the qubit state and leaves subsequent measurements consistent with the new state, producing an isolated $S_1$.
An assignment error changes only one reported bit and therefore breaks the correlations on both sides of that readout, producing $S_2$.
Long-lived leakage can break several consecutive correlations and appears as $S_{3+}$. Short-lived leakage does not generally persist long enough to form $S_{3+}$ and is instead detected through the excess $S_1$ under an $X$-$\tau$-$X$ sequence.
Figure~\ref{fig:error_taxonomy}(d) shows measured examples of the three syndrome classes, with the corresponding examples of repeated-readout IQ traces in Supplementary Fig.~\ref{fig:supp:qnd_traces}.

RED directly counts the number of error events and normalizes by the effective number of readouts to obtain the corresponding error rates.
The syndrome--channel correspondences define the RED decoder summarized in Table~\ref{tab:decoder}, while the five measurement sequences used to construct the complete decoder are listed in Table~\ref{tab:red_experiments}.
\begingroup
\squeezetable
\begin{table}[!htb]
\caption{RED decoder dictionary.
Each entry is written as $\mathrm{sequence}_{\mathrm{branch}}(\mathrm{syndrome})$:
$I$, $X$, and $X$-$\tau$-$X$ label the inter-readout gate sequence used in that experiment, the subscript denotes the prepared branch, and the parentheses indicate the syndrome class extracted from that run.
Short-lived leakage is estimated from the excess $S_1$ of an $X$-$\tau$-$X$ sequence relative to the matched $I$ baseline.}
\label{tab:decoder}
\resizebox{\linewidth}{!}{%
\begin{tabular}{lcccc}
\toprule
Branch &
Bit-flip &
Assignment &
Short leak. &
Long leak. \\
\midrule
$\ket{g}$ &
$I_{\ket{g}}(S_1)$ &
$I_{\ket{g}}(S_2)$ &
$(X\!-\!\tau\!-\!X)_{\ket{g}}(S_1)\!-\!I_{\ket{g}}(S_1)$ &
$X_{\ket{g}\text{-parity}}(S_{3+})$ \\
$\ket{e}$ &
$I_{\ket{e}}(S_1)$ &
$I_{\ket{e}}(S_2)$ &
$(X\!-\!\tau\!-\!X)_{\ket{e}}(S_1)\!-\!I_{\ket{e}}(S_1)$ &
$X_{\ket{e}\text{-parity}}(S_{3+})$ \\
\bottomrule
\end{tabular}}%
\end{table}
\endgroup

\begingroup
\squeezetable
\begin{table}[!htb]
\caption{Five RED experiments used to populate the decoder dictionary.}
\label{tab:red_experiments}
\begin{ruledtabular}
\begin{tabular}{llll}
Exp. & Init. & Gate & Decoder usage \\
\hline
$I_{\ket{g}}$ & $\ket{g}$ & $I$ & bit-flip \& assignment $\ket{g}$ \\
$I_{\ket{e}}$ & $\ket{e}$ & $I$ & bit-flip \& assignment $\ket{e}$ \\
$X$ & $\ket{g}/\ket{e}$ & $X$ & long-lived leakage \\
$(X\!-\!\tau\!-\!X)_{\ket{g}}$ & $\ket{g}$ & $X$-$\tau$-$X$ & short-lived leakage $\ket{g}$ \\
$(X\!-\!\tau\!-\!X)_{\ket{e}}$ & $\ket{e}$ & $X$-$\tau$-$X$ & short-lived leakage $\ket{e}$ \\
\end{tabular}
\end{ruledtabular}
\end{table}
\endgroup

In the remainder of this paper, we use three devices, two fluxonium qubits (Fluxonium 1 and Fluxonium 2) and a transmon, summarized in \tableautorefname\;\ref{tab:fig4_summary}, to validate RED and demonstrate how it separates and quantifies different readout-error mechanisms. The main-text figures focus on Fluxonium 1 and the Transmon, while complementary results for Fluxonium 2 are presented in Appendix~\ref{app:fluxonium2}.

\subsection{Validation of RED error extraction}
\label{sec:results:validation}

We first validate the channel specificity of RED through controlled bit-flip and leakage injections.

\begin{figure}[t]
  \includegraphics[width=\linewidth]{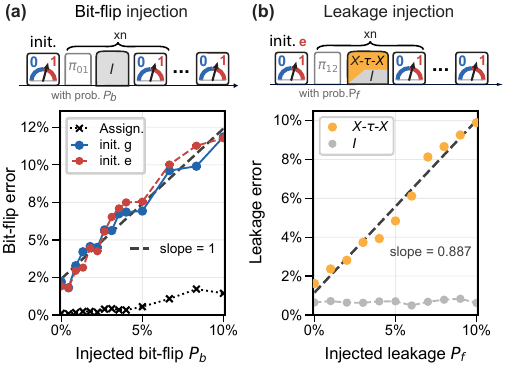}
  \caption{\label{fig:validation}\textbf{Controlled error-injection validation of RED.}
  \textbf{(a)}~Bit flips are injected by a $\pi_{01}$ pulse applied with probability $P_b$ on $I$ sequences prepared in $\ket{g}$ and $\ket{e}$: the extracted bit-flip channel tracks $P_b$ with near-unit slope on both preparations, while the assignment channel remains nearly flat.
  \textbf{(b)} Leakage is injected by a $\pi_{12}$ pulse applied with probability $P_f$ starting from $\ket{e}$. The extracted leakage error from the $X$-$\tau$-$X$ sequence increases linearly with $P_f$ with a fitted slope of 0.887, while the raw $S_1$ rate of the $I$ reference remains nearly constant. Data are taken on fluxonium qubits.}
\end{figure}

Figure~\ref{fig:validation}(a) shows the bit-flip injection experiment. A $\pi_{01}$ pulse is applied with a programmed probability $P_b$ in each cycle of an all-$I$ repeated-readout sequence, with the qubit initialized in either $\ket{g}$ or $\ket{e}$. For both preparation branches, the extracted bit-flip rate increases approximately linearly with $P_b$, with a slope close to unity. In contrast, the assignment-error channel remains nearly unchanged, showing only a slight increase at higher injection probabilities. This residual $S_2$ increase is consistent with two bit flips occurring in adjacent cycles, which produce two consecutive broken correlations and can therefore mimic an assignment error (Appendix \ref{app:red_recovery}).

We next validate the leakage channel, as shown in Fig.~\ref{fig:validation}(b). Initialized in $\ket{e}$, we apply a $\pi_{12}$ pulse with a programmed probability $P_f$ to inject population into the leakage $\ket{f}$
state. The extracted leakage error increases approximately linearly with $P_f$, while the $I$ reference remains nearly constant. This contrast demonstrates the need for inter-readout $X$ pulses to detect these leakage events.
The fitted slope is 0.887, slightly below unity. This is consistent with incomplete $\ket{e}\rightarrow\ket{f}$ population transfer due to the short lifetime of $\ket{f}$. The detailed leakage extraction protocol is discussed in the following subsection.

\begin{figure}[!t]
  \includegraphics[width=\linewidth]{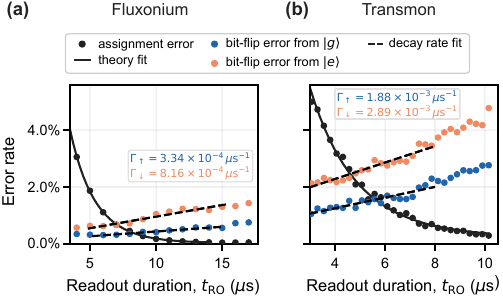}
  \caption{\label{fig:conventional} 
  \textbf{Assignment and bit-flip errors versus readout duration.} \textbf{(a)}~Fluxonium and \textbf{(b)}~transmon. 
  The assignment-error rate decreases with increasing $t_{\RO}$ as the readout SNR improves, while the bit-flip rates from $\ket{g}$ and $\ket{e}$ increase approximately linearly over the fitted range. Solid black curves show Gaussian-overlap fits to the assignment errors, while dashed lines show linear fits to the bit-flip rates, yielding the excitation and relaxation rates $\Gamma_\uparrow$ and $\Gamma_\downarrow$. The fitting models are detailed in Appendix~\ref{app:fit_models}.}
\end{figure}

We next apply RED to a fluxonium and a transmon, varying the readout duration in Fig.~\ref{fig:conventional} to examine the assignment and bit-flip error channels. On both devices, the extracted assignment-error rate decreases rapidly with integration time, as the increasing signal-to-noise ratio (SNR) reduces the overlap between the readout signal distributions of $\ket{g}$ and $\ket{e}$. The observed dependence agrees well with a Gaussian-overlap model within the fitted range, supporting the identification of $S_2$ syndromes with assignment errors. The fitting procedure is detailed in Appendix~\ref{app:fit_models}.

Bit-flip errors can occur during the readout pulse and therefore the error rate is expected to increase with the readout duration \(t_{\RO}\). In the absence of additional readout-induced processes, this increase is governed by the device's intrinsic relaxation  \(T_1\) and thermal-excitation rates within the computational subspace.
In Fig.~\ref{fig:conventional}, the branch-resolved bit-flip channels grow approximately linearly at intermediate durations.
We therefore fit the linear model of Appendix~\ref{app:fit_models} to extract the excitation and relaxation rates $\Gamma_{\uparrow}$ and $\Gamma_{\downarrow}$, or equivalently
$T_{\uparrow/\downarrow}=1/\Gamma_{\uparrow/\downarrow}=2990/1230\,\mu\mathrm{s}$ for Fluxonium
and $532/346\,\mu\mathrm{s}$ for the transmon.
Combining the two branches gives
$T_q=(1/T_{\uparrow}+1/T_{\downarrow})^{-1}=872\,\mu\mathrm{s}$ (Fluxonium) and $210\,\mu\mathrm{s}$ (transmon).
These values are to be compared with the independently measured $T_1=2130\pm360\,\mu\mathrm{s}$ and $274\pm30\,\mu\mathrm{s}$ (Appendix~\ref{app:device_params}).
In both devices, $T_q$ is shorter than the independently measured $T_1$ \cite{Huang2026Readout}. The mismatch is not surprising. Under driven readout, bit flips need not follow idle $T_1$ alone; the measurement tone can open additional bit-flip error channels, for example second-order transitions via higher states~\cite{dumas2024ionization,shillito2022,sank2016mist}.
Together, the error-injection experiments and the readout-duration sweeps show that RED separates assignment from bit-flip errors and recovers both rates.

\subsection{Leakage error extraction}
\label{sec:results:leakage}
\begin{figure*}[t]
  \centering
  \includegraphics[width=\textwidth]{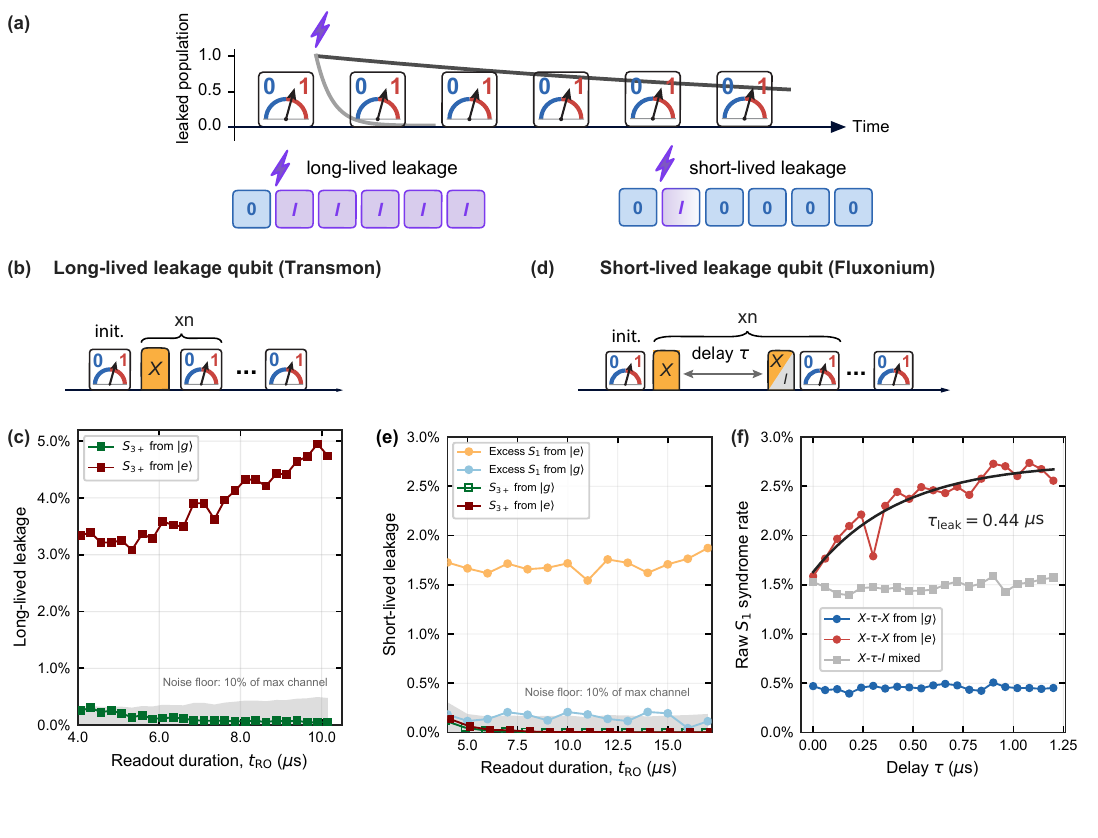}
  \caption{\textbf{RED leakage detection across long- and short-lived regimes.}
  \textbf{(a)}~Leaked population during repeated readouts. A long-lived leakage persists over many readout cycles and breaks a run of correlations; a short-lived leakage returns within one cycle and  breaks only one correlation at most. 
  \textbf{(b)}~Long-lived-leakage probe: repeated readouts with an inter-readout $X$ gate.
  \textbf{(c)}~Transmon data, $S_{3+}$ versus readout duration.
  The rate from $\ket{e}$ rises with $t_{\RO}$; the rate from $\ket{g}$ stays at the noise floor. 
  The gray band marks a reference threshold equal to $10\%$ of the largest decoded error channel. The same band is shown in (e).
  \textbf{(d)}~Short-lived-leakage probe: an $X-\tau-X$ sequence between successive readouts.
  \textbf{(e)}~Fluxonium data: excess $S_1^{X\text{-}\tau\text{-}X}-S_1^{I}$ and $S_{3+}$ versus readout duration, measured at a fixed delay $\tau=1.5\,\mu\mathrm{s}$.
  The excess $S_1$ from $\ket{e}$ remains nearly constant with $t_{\RO}$, whereas that from $\ket{g}$ stays near the noise floor.
  The $S_{3+}$ rates from both branches also remain at the noise floor, showing that this short-lived leakage is revealed by the excess $S_1$ but not by $S_{3+}$.
  \textbf{(f)}~Raw $S_1$ rates versus the inter-pulse delay $\tau$. For the $X$-$\tau$-$X$ sequence, the rate from $\ket{e}$ rises and saturates, yielding a leakage lifetime $\tau_{\mathrm{leak}}=0.44\,\mu\mathrm{s}$ from an exponential fit.} 
  \label{fig:leakage}

  \vspace{0.6em}
  \captionof{table}{\label{tab:fig4_summary}Decoded RED error channels and standard benchmarks at the readout duration that minimizes the total RED error for each device.
  Branch-resolved columns list the $\ket{e}$ and $\ket{g}$ contributions separately; a dash means that leakage channel is not applicable.
  Measurement infidelity ($1-F_{\mathrm{m}}$) and QNDness error ($1-R$) are standard benchmarks at the same operating point.
  The Total column is the sum of the assignment error and the branch-averaged bit-flip and leakage rates.}
 
  \begin{ruledtabular}
  \begin{tabular}{lcccccc|cc}
    & &
    \multicolumn{5}{c}{RED error channels (\%)} &
    \multicolumn{2}{c}{Standard benchmarks (\%)} \\
    \cline{3-7}\cline{8-9}
    Device &
    $t_{\RO}$ ($\mu$s) &
    Assignment &
    Bit-flip &
    Short-lived leakage &
    Long-lived leakage &
    Total &
    Meas. infidelity &
    QNDness error \\
    \hline
    Fluxonium 1 &
    11 &
    0.11 &
    $e$: 1.13; $g$: 0.46 &
    $e$: 1.54; $g$: 0.23 &
    -- &
    1.79 &
    0.28 &
    0.72 \\
    Fluxonium 2 &
    6.0 &
    0.10 &
    $e$: 1.94; $g$: 1.27 &
    $e$: 0.03; $g$: 0.31 &
    -- &
    1.88 &
    0.53 &
    2.05 \\
    Transmon &
    6.5 &
    1.10 &
    $e$: 2.90; $g$: 1.51 &
    -- &
    $e$: 3.49; $g$: 0.12 &
    5.11 &
    5.43 &
    4.41 \\
  \end{tabular}
  \end{ruledtabular}
\end{figure*}

Having established the assignment and bit-flip channels, we next turn to leakage.
Leakage is harder to diagnose from binary records than the previous two channels.
The leaked population may occupy different higher states, which are generally unknown a priori and are not resolved by a two-state discriminator in the IQ plane~\cite{sank2016mist,dumas2024ionization,wang2026probing}.
The lifetime of that population can also be long or short compared with the readout cycle, as shown in Fig.~\ref{fig:leakage}(a).

When $\tau_{\mathrm{leak}}\gg t_{\mathrm{cycle}}$, the leaked population persists across successive readouts.
An $X$ gate between readouts converts this into a run of broken correlations, which we classify as $S_{3+}$, as shown in Fig.~\ref{fig:leakage}(b). The X gate sequence in the absence of readout errors switches the state of the qubit before a readout is done in every cycle. Therefore, half of the readouts in a string are performed with the qubit in $|g\rangle$ and the other half in $|e\rangle$. The total error recorded with this sequence is therefore an average of the two branch-resolved errors. The contributions can be separated by keeping track of the parity of the position at which the error is recorded.  If we assume that in the beginning of the protocol the qubit is in the ground state,  all $S_{3+}$ errors recorded at odd positions will originate from the qubit leaking outside of the computational subspace while the readout was performed on $|e\rangle$ state and vice-versa.

When $\tau_{\mathrm{leak}}\lesssim t_{\mathrm{cycle}}$, the population can return to the computational subspace before the next readout, so it leaves no persistent syndrome and may not even appear as a distinct outcome in a single binary shot.
If the leakage happens to a short lived state that decays to the opposite computational state the error will be recorded as $S_1$ syndrome and we will classify this as a bit-flip error, therefore such an event will be captured by the decoder. However, in the case where the readout causes  leakage to a short lived state which decays to the same state, the error is not detected directly. This event can still be recorded with the use of inter-readout gate sequence. 

For this purpose we use two $\pi_{01}$ pulses separated by a delay $\tau$, as shown in Fig.~\ref{fig:leakage}(d). In this case if there was a leakage event to a short lived state, the first $\pi_{01}$ pulse in this sequence will fail, then if the leaked state decays to the same state as the initial one during the delay $\tau$, the second pulse will flip the qubit state and the error would be detected as $S_1$ syndrome. On the other hand, if leaked state does not decay back to the computational subspace during the wait $\tau$ the second pulse also fails and the error will not be recorded since the leaked state decays to the initial state during the readout. 
For this reason the delay between the two pulses is essential for the error detection and the wait time $\tau$ should be several times the leakage lifetime in order to fully capture the short-lived leakage error.

The two devices discussed in Fig.~\ref{fig:leakage} show these two lifetime regimes: long-lived leakage on the transmon and short-lived leakage on the fluxonium.

The transmon data match the long-lived case: leaked population persists across successive readouts and appears as $S_{3+}$ on the $X$ sequence. On this transmon, the extracted leakage occurs predominantly from $\ket{e}$.
The leakage rate from $\ket{e}$ increases with readout duration, as shown in Fig.~\ref{fig:leakage}(c), while the $\ket{g}$ contribution remains below the reference threshold. Within our resolution, readout-induced leakage from $\ket{g}$ is therefore negligible.
These rates are the long-lived leakage entries in Table~\ref{tab:fig4_summary}.

The fluxonium data match the short-lived regime.
The $S_{3+}$ rate remains far below the noise floor in Fig.~\ref{fig:leakage}(e), so $S_{3+}$ cannot be used to extract the short-lived leakage rate.
This is expected because the leakage lifetime is shorter than the readout cycle: the leaked population does not persist across the record and therefore does not form $S_{3+}$.
To probe the leakage state before it returns to the computational subspace, we insert a delay between two $\pi_{01}$ pulses ($X$-$\tau$-$X$), as shown in Fig.~\ref{fig:leakage}(d).
The resulting mismatch appears as a single broken correlation, i.e.\ an $S_1$ event.
Because an $I$ sequence measures the ordinary bit-flip contribution to $S_1$, we subtract this baseline; the excess
$S_1^{X\text{-}\tau\text{-}X}-S_1^{I}$ isolates the short-lived leakage contribution.

As shown in Fig.~\ref{fig:leakage}(e), the extracted leakage on this fluxonium occurs predominantly from $\ket{e}$, while the $\ket{g}$ contribution remains almost at the noise floor.
This branch dependence differs on
Fluxonium 2: leakage from $\ket{g}$ and from $\ket{e}$ is more nearly balanced, as shown in Supplementary Fig.~\ref{fig:supp:fluxonium2}.
Readout-induced leakage therefore depends on the qubit and readout parameters.
A detailed study of these microscopic dependencies is beyond the scope of this paper and will be reported separately.
The excess-$S_1$ values provide the fluxonium short-lived leakage entry in Table~\ref{tab:fig4_summary}.

We next vary the inter-pulse delay in Fig.~\ref{fig:leakage}(f).
During the delay $\tau$, leaked population can return to the computational subspace before the second $\pi_{01}$ pulse.
Sweeping $\tau$ therefore changes how much leakage is still present at the second pulse, and $S_1$ versus $\tau$ measures the leakage lifetime.
The $\ket{e}$-branch $X$-$\tau$-$X$ response grows with $\tau$ and is well described by a single exponential with $\tau_{\mathrm{leak}}=0.44\,\mu\mathrm{s}$.
By contrast, the raw $S_1$ rate on the $\ket{g}$ branch is almost independent of $\tau$, indicating that this signal is dominated by bit-flip transitions and that short-lived leakage from $\ket{g}$ is negligible.
The $X$-$\tau$-$I$ sequence is a control experiment. With only one $X$ pulse, the qubit alternates between $\ket{g}$ and $\ket{e}$, so the measured short-lived leakage is an average of the two branches. 
We note that the short-lived leakage observed here is not necessarily a generic property of fluxonium. In the two fluxonium devices, leakage may involve the $\ket{2}$ or $\ket{3}$ states, whose transition frequencies lie close to the readout resonator frequency, leading to a strong Purcell-induced relaxation. The short leakage lifetime may therefore be specific to the level structure and resonator coupling of these devices.

To sum up, RED covers both lifetime regimes with the same decoder: long-lived leakage appears as $S_{3+}$ on the $X$ sequence, while short-lived leakage is read as excess $S_1$ under $X$-$\tau$-$X$.
In principle, an $X$-$\tau$-$X$ delay sweep extracts the leakage lifetime on any qubit that supports this sequence, independent of the qubit modality.

\subsection{Summary of RED error channels}
\label{sec:results:error_budget}

Table~\ref{tab:fig4_summary} summarizes the channel-resolved RED error budgets for all three devices at the readout duration that minimizes the total RED error. We report assignment, bit-flip, short-lived-leakage, and long-lived-leakage errors, with the latter three resolved by the $\ket{g}$ and $\ket{e}$ branches.

Fluxonium 1 and Fluxonium 2 have similar total RED errors, but their channel compositions differ. On Fluxonium 1, bit-flip and short-lived-leakage errors
contribute comparably, whereas bit flips dominate on Fluxonium 2. The branch-resolved leakage contributions also differ between the two devices. The transmon has a larger total error and additionally exhibits long-lived leakage. Thus, similar total error rates can arise from distinct underlying mechanisms.

For comparison, Table~\ref{tab:fig4_summary} also lists the measurement infidelity and QNDness error at the same operating points.
Measurement infidelity quantifies errors in discriminating $\ket{g}$ and $\ket{e}$, but does not separately resolve leakage. QNDness error is sensitive to changes between repeated outcomes, but leakage may remain hidden if the binary discriminator maps the leaked population onto the same outcome~\cite{rilb}.

By separating the total error into physical error channels and resolving their $\ket{g}$ and $\ket{e}$ contributions, RED reveals the dominant readout-error mechanism on each device---information that is not available from Measurement infidelity or QNDness alone.

\section{Discussion}
\label{sec:discussion}

RED decodes repeated binary readout records into a channel-resolved error budget for assignment, bit-flip, and leakage errors, resolved by the $\ket{g}$ and $\ket{e}$ branches.
The extracted channels are validated by controlled error injection and by their expected dependence on readout duration.
Across the devices studied here, RED identifies error mechanisms that conventional measurement fidelity or repeatability metrics cannot
distinguish. In particular, devices with similar total error rates can exhibit substantially different error-channel compositions.

A key advantage of RED is its ability to characterize leakage without directly resolving non-computational states in the IQ plane. We demonstrate RED on both fluxonium and transmon devices. By selecting appropriate inter-readout gate sequences, RED detects both
long- and short-lived leakage using only thresholded binary  measurements. The protocol therefore does not require the leakage population to persist across successive readouts. This allows the same decoding framework to be applied across qubit devices with different leakage dynamics.

The syndrome-to-channel mapping nevertheless has limitations. Different error processes can produce the same syndrome, while overlapping errors introduce higher-order contamination. This ambiguity is intrinsic to syndrome-based inference and also arises in quantum error correction~\cite{dennis2002,fowler2012}. We quantify these effects for assignment and bit-flip errors using simulated repeated-readout records with known error rates (Supplementary Fig.~\ref{fig:supp:red_error_maps}). The simulations show that the dominant channel is recovered accurately over a broad parameter range, whereas the absolute accuracy of the smaller channel degrades when the error rates are strongly imbalanced. Even in this regime, the leading error channel and the qualitative error hierarchy remain informative.

Beyond superconducting qubits, the same analysis could in principle be adapted to other qubit platforms supporting repetitive QND measurement, with the syndrome definitions tailored to their relevant error processes~\cite{xue2020prx,yoneda2020qnd,hume2007qnd,negnevitsky2018,graham2023mcm,norcia2023mcm,zeng2026fndr}.
A natural next step is to extend RED to faster readout. The measurements reported here do not use parametric amplification, and the readout durations are therefore typically several microseconds. With parametric amplification, reducing the readout time to the sub-microsecond regime would provide finer resolution of how the individual error channels develop during measurement. Combined with sweeps of readout power and frequency, these channel-resolved dynamics may help constrain the microscopic origin of readout-induced errors, including intrinsic multi-photon transitions~\cite{dumas2024ionization} and processes involving two-level systems or spurious electromagnetic modes~\cite{dai2025dust}. RED could therefore provide not only an error budget, but also a tool for investigating the physical mechanisms that limit QND readout.

\section*{Data and Code Availability}

The RED (Readout Error Decoder) Python package, including the decoder implementation, tutorials, and raw IQ example datasets from the fluxonium and transmon measurements presented in this work, is publicly available
on Zenodo~\cite{zheng2026red}. Additional data and code are available from the authors upon reasonable request.

\section*{Author contributions}
H.Z.\ and V.P.\ contributed equally to this work.
They proposed the RED protocol, performed the measurements, and carried out the data analysis.
N.G.\ contributed to the measurements and data analysis.
A.L.\ performed the numerical simulations of Supplementary Fig.~\ref{fig:supp:red_error_maps}.
T.I.\ and C.A.\ acquired the transmon data.
M.N.M.\ fabricated the devices.
R.M.\ contributed to the installation of the measurement lines.
H.Z.\ wrote the manuscript with input from all authors.
E.C.\ and V.E.M.\ supervised the project.

\begin{acknowledgments}

We thank Andrea Duina for assistance with coding and developing the RILB analysis framework, and Chien-An Wang and Nikola Dimitrov for fruitful discussions.
This work was supported by the Swiss National Science Foundation (SNSF) under Grant No. 514944; NCCR SPIN, a National Centre of Competence in Research funded by the SNSF under Grant No. 225153; the Swiss State Secretariat for Education, Research and Innovation (SERI) under Grant No. 589078; and the U.S. Army Research Office (ARO) through the QCS5 program under Grant No. W911NF-23-1-0093.

\end{acknowledgments}

\appendix

\twocolumngrid

\setcounter{figure}{0}
\renewcommand{\thefigure}{S\arabic{figure}}
\renewcommand{\theHfigure}{S.\arabic{figure}}

\section{RED workflow and error-rate extraction}
\label{app:red_extraction}

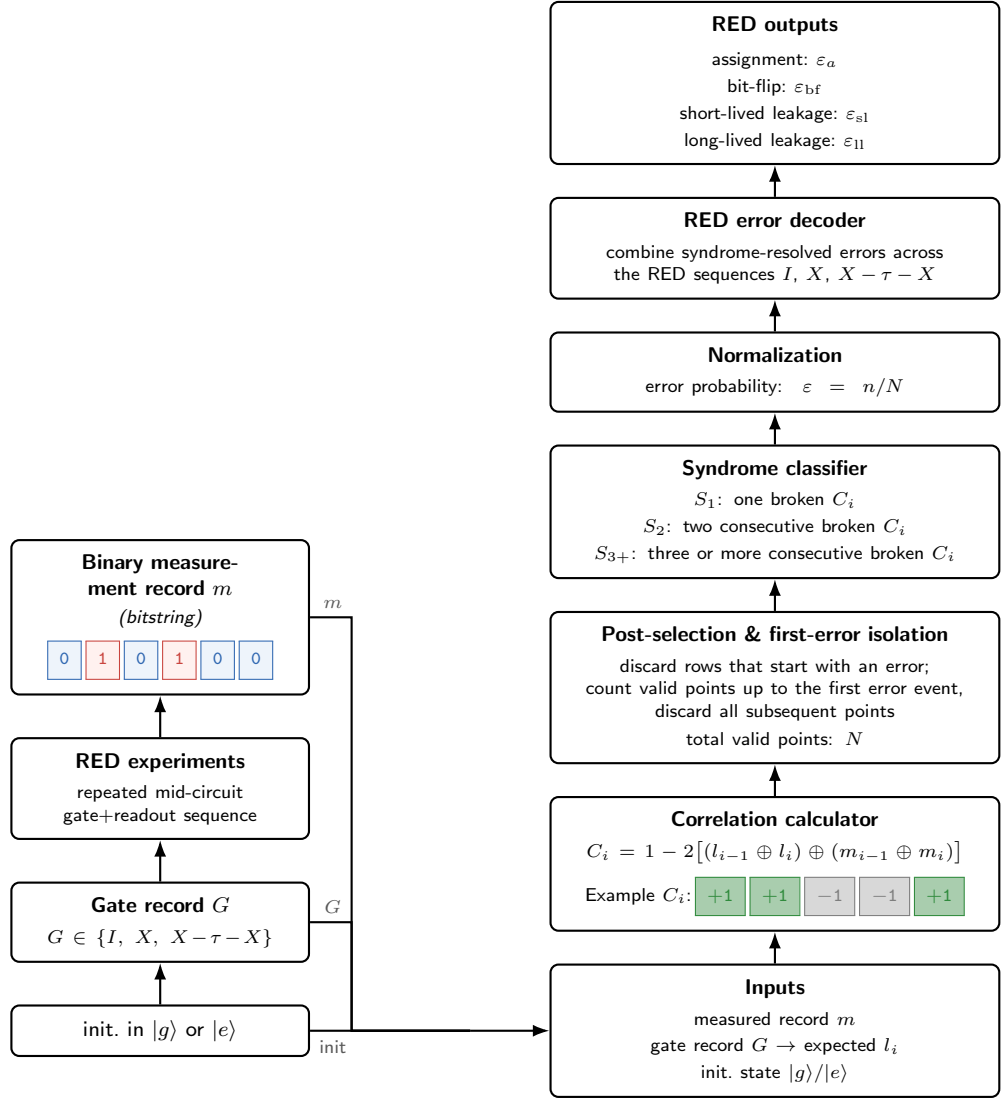
\begin{figure*}[t]
  \centering
\definecolor{FigBlue}{HTML}{2F67B1}
\definecolor{FigBlueFill}{HTML}{EEF4FB}
\definecolor{FigRed}{HTML}{C9443E}
\definecolor{FigRedFill}{HTML}{FFF1EF}
\definecolor{FigGreen}{HTML}{2F8A3C}
\definecolor{FigCPos}{HTML}{A9CDAE}
\definecolor{FigCNeg}{HTML}{D8D8D8}
\definecolor{FigCNegEdge}{HTML}{7D7D7D}

\providecommand{\figchip}[3]{%
  {\color{#1}%
    \fcolorbox{#1}{#2}{\scriptsize\ttfamily\bfseries\strut\color{#1}\,#3\,}}%
}
\newcommand{\bitzero}{\figchip{FigBlue}{FigBlueFill}{0}}
\newcommand{\bitone}{\figchip{FigRed}{FigRedFill}{1}}
\newcommand{\cplus}{\figchip{FigGreen}{FigCPos}{$+$1}}
\newcommand{\cminus}{\figchip{FigCNegEdge}{FigCNeg}{$-$1}}

\begin{tikzpicture}[
  font=\footnotesize\sffamily,
  >=Latex,
  node distance=0.42cm and 1.55cm,
  every node/.style={align=center},
  inputbox/.style={
    draw=black,
    thick,
    rounded corners=3pt,
    inner sep=5.5pt,
    text width=3.55cm,
  },
  experimentbox/.style={
    draw=black,
    thick,
    rounded corners=3pt,
    inner sep=5.5pt,
    text width=3.55cm,
  },
  processingbox/.style={
    draw=black,
    thick,
    rounded corners=3pt,
    inner sep=5.5pt,
    text width=5.55cm,
  },
  peachproc/.style={
    draw=black,
    thick,
    rounded corners=3pt,
    inner sep=5.5pt,
    text width=5.55cm,
  },
  decoderbox/.style={
    draw=black,
    thick,
    rounded corners=3pt,
    inner sep=5.5pt,
    text width=5.55cm,
  },
  outputbox/.style={
    draw=black,
    thick,
    rounded corners=3pt,
    inner sep=5.5pt,
    text width=5.55cm,
  },
  flowarrow/.style={
    -{Latex[length=2.2mm,width=1.6mm]},
    thick,
    black,
  },
]

\node[decoderbox] (R1) {%
  \textbf{Inputs}\\[2pt]
  {\scriptsize measured record $m$}\\[0.5pt]
  {\scriptsize gate record $G$ $\rightarrow$ expected $l_i$}\\[0.5pt]
  {\scriptsize init.\ state $\ket{g}$/$\ket{e}$}%
};

\node[peachproc, above=of R1] (R2) {%
  \textbf{Correlation calculator}\\[2pt]
  {\scriptsize
  $C_i=1-2\bigl[(l_{i-1}\oplus l_i)\oplus(m_{i-1}\oplus m_i)\bigr]$}\\[4pt]
  {\scriptsize Example $C_i$:\,}%
  \cplus\,\cplus\,\cminus\,\cminus\,\cplus%
};

\node[processingbox, above=of R2] (R3) {%
  \textbf{Post-selection \& first-error isolation}\\[2pt]
  {\scriptsize discard rows that start with an error;}\\[-1pt]
  {\scriptsize count valid points up to the first error event,}\\[-1pt]
  {\scriptsize discard all subsequent points}\\[1pt]
  {\scriptsize total valid points: $N$}%
};

\node[processingbox, above=of R3] (R5) {%
  \textbf{Syndrome classifier}\\[2pt]
  {\scriptsize $S_1$: one broken $C_i$}\\[0.5pt]
  {\scriptsize $S_2$: two consecutive broken $C_i$}\\[0.5pt]
  {\scriptsize $S_{3+}$: three or more consecutive broken $C_i$}%
};

\node[processingbox, above=of R5] (R6) {%
  \textbf{Normalization}\\[2pt]
  {\scriptsize error probability:\quad $\varepsilon = n/N$}%
};

\node[decoderbox, above=of R6] (R7) {%
  \textbf{RED error decoder}\\[2pt]
  {\scriptsize combine syndrome-resolved errors across}\\[-1pt]
  {\scriptsize the RED sequences $I$, $X$, $X\!-\!\tau\!-\!X$}%
};

\node[outputbox, above=of R7] (R8) {%
  \textbf{RED outputs}\\[3pt]
  {\scriptsize assignment:\ $\varepsilon_a$}\\[0.7pt]
  {\scriptsize bit-flip:\ $\varepsilon_{\mathrm{bf}}$}\\[0.7pt]
  {\scriptsize short-lived leakage:\ $\varepsilon_{\mathrm{sl}}$}\\[0.7pt]
  {\scriptsize long-lived leakage:\ $\varepsilon_{\mathrm{ll}}$}%
};

\draw[flowarrow] (R1) -- (R2);
\draw[flowarrow] (R2) -- (R3);
\draw[flowarrow] (R3) -- (R5);
\draw[flowarrow] (R5) -- (R6);
\draw[flowarrow] (R6) -- (R7);
\draw[flowarrow] (R7) -- (R8);

\node[inputbox] (L1) at ([xshift=-5.15cm]R1.west) {%
  init.\ in $\ket{g}$ or $\ket{e}$%
};

\node[inputbox, above=0.55cm of L1] (L2) {%
  \textbf{Gate record $G$}\\[2pt]
  {\scriptsize $G\in\{I,\ X,\ X\!-\!\tau\!-\!X\}$}%
};

\node[experimentbox, above=0.55cm of L2] (L3) {%
  \textbf{RED experiments}\\[2pt]
  {\scriptsize repeated mid-circuit}\\[-1pt]
  {\scriptsize gate+readout sequence}%
};

\node[inputbox, above=0.55cm of L3] (L4) {%
  \textbf{Binary measurement record $m$}\\[1pt]
  {\scriptsize\textit{(bitstring)}}\\[4pt]
  \bitzero\,\bitone\,\bitzero\,\bitone\,\bitzero\,\bitzero%
};

\draw[flowarrow] (L1) -- (L2);
\draw[flowarrow] (L2) -- (L3);
\draw[flowarrow] (L3) -- (L4);

\coordinate (junc) at ($(R1.west)+(-1.05,0)$);
\draw[thick, black]
  (L4.east) -- ++(0.55,0)
    node[pos=0.55, above, font=\scriptsize\sffamily, text=black!65] {$m$}
    |- (junc);
\draw[thick, black]
  (L2.east) -- ++(0.55,0)
    node[pos=0.55, above, font=\scriptsize\sffamily, text=black!65] {$G$}
    |- (junc);
\draw[thick, black]
  (L1.east) -- ++(0.55,0)
    node[pos=0.55, below, font=\scriptsize\sffamily, text=black!65] {init}
    |- (junc);
\draw[flowarrow] (junc) -- (R1.west);

\end{tikzpicture}
  \caption{\textbf{RED workflow: binary single-shot records to error probabilities.}
  \textbf{Left:} preparation in $\ket{g}$ or $\ket{e}$, the gate record $G$ with $G\in\{I,X,X\text{-}\tau\text{-}X\}$ fixed within each experiment, and the repeated gate--readout sequence produce the binary measurement record $m$.
  \textbf{Right:} the measured record $m$, gate record $G$, and initial state determine the expected logical record $l_i$ and the correlation record
  $C_i = 1 - 2[(l_{i-1}\oplus l_i)\oplus(m_{i-1}\oplus m_i)]$.
  The first error event in each record is isolated and classified as $S_1$, $S_2$, or $S_{3+}$. Syndrome counts are normalized by the effective readout count $N$, and the decoder dictionary converts the resulting syndrome rates into assignment ($\varepsilon_a$), bit-flip ($\varepsilon_{\mathrm{bf}}$), short-lived leakage ($\varepsilon_{\mathrm{sl}}$), and long-lived leakage ($\varepsilon_{\mathrm{ll}}$) channel estimators.}
  \label{fig:supp:red_workflow}
\end{figure*}

This section describes the complete RED workflow used to convert repeated binary readout records into channel-resolved error rates. Supplementary Fig.~\ref{fig:supp:red_workflow} summarizes the procedure.

For most of the data presented in this work, each RED measurement
sequence consists of 50 consecutive readouts and is repeated
500 times, separately for the $\ket{g}$ and $\ket{e}$
preparation branches.
The syndrome counts are accumulated over all repetitions
to extract the error rates as follows:

\begin{enumerate}

  \item \textbf{Construct the expected and measured records.}
  The prepared state and the fixed inter-readout operation $G$ define the expected logical record $l_i$. Thresholding each single-shot readout gives the measured binary record $m_i$.

  \item \textbf{Construct the correlation record.}
  The expected and measured changes between neighboring readouts are compared through
  \begin{equation}
  C_i =
  1-2\bigl[(l_{i-1}\oplus l_i)\oplus(m_{i-1}\oplus m_i)\bigr].
  \end{equation}
  Here $C_i=+1$ denotes agreement with the expected evolution and $C_i=-1$ denotes a broken correlation.

  \item \textbf{Isolate the first error event.}
  Each record is scanned until the first broken correlation is encountered. All data following the first error event are excluded from the error-rate extraction so that each retained record contributes at most one error event. Records that begin with an error are discarded, as they typically indicate initialization errors.

  \item \textbf{Classify the first-error syndrome.}
  Starting from the first broken correlation, consecutive values with $C_i=-1$ are grouped by run length:
  \[
  S_1:\;1,\qquad
  S_2:\;2,\qquad
  S_{3+}:\;\geq3
  \]
  consecutive broken correlations. The corresponding numbers of first-error events are denoted by $n_{S_1}$, $n_{S_2}$, and $n_{S_{3+}}$.

  \item \textbf{Determine the effective readout count.}
  For each retained record, only the correct readouts preceding the first error contribute to the exposure count. Summing these valid readouts over all records defines the effective count $N$.

  \item \textbf{Normalize the syndrome counts.}
  The syndrome rates are obtained directly from
  \begin{equation}
  \varepsilon_{S_j}=\frac{n_{S_j}}{N},
  \qquad
  j\in\{1,2,3+\}.
  \end{equation}

  \item \textbf{Decode the physical error channels.}
  The syndrome rates from the $I$, $X$, and $X$-$\tau$-$X$ experiments are combined according to the decoder dictionary in Table~\ref{tab:decoder}. This gives the assignment, bit-flip, short-lived-leakage, and long-lived-leakage estimators, resolved by the $\ket{g}$ and $\ket{e}$ branches where applicable.

\end{enumerate}

For the all-$X$ sequence in Fig.\ref{fig:leakage}(c), the $\ket{g}$ and $\ket{e}$ branches each contributes approximately half of the valid readouts, so branch-resolved $S_{3+}$ rates are normalized by $N/2$.

\section{QND traces of the three readout errors}
\label{app:qnd_traces}

\begin{figure*}[t]
  \centering
  \includegraphics[width=\textwidth]{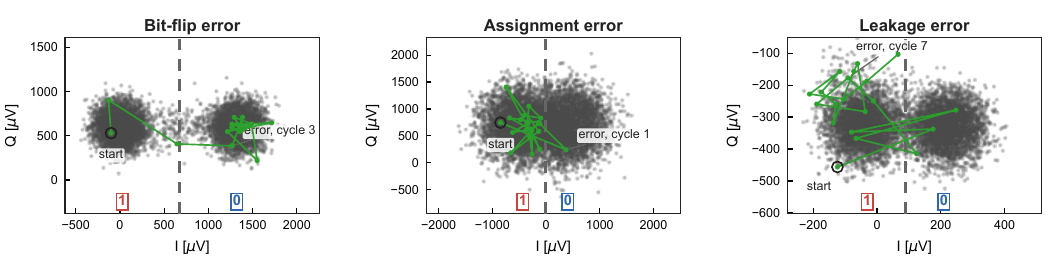}
  \caption{QND IQ traces for Fig.~\ref{fig:error_taxonomy}(d): bit-flip, assignment, and leakage.}
  \label{fig:supp:qnd_traces}
\end{figure*}

Supplementary Fig.~\ref{fig:supp:qnd_traces} shows one repeated-readout IQ trace for each syndrome example in Fig.~\ref{fig:error_taxonomy}(d).

\section{Error-injection protocol}
\label{app:error_injection}

For the validation in Fig.~\ref{fig:validation}, each measurement sequence consists of n=100 consecutive readouts. For each programmed injection probability $P_b$ or $P_f$, we use at least 20 independently generated random masks specifying the injection-pulse positions within the sequence, with at least 500 shots acquired per mask. The syndrome counts are accumulated over all masks and shots to extract the error rates. The $\pi_{01}$ bit-flip injection uses all-$I$ sequences prepared in $\ket{g}$ and $\ket{e}$, while the $\pi_{12}$ leakage injection uses the $X$-$\tau$-$X$ sequence with an $I$ reference.

The average fidelity of the $\pi_{01}$ pulses exceeds 99.8\%. The $\pi_{12}$ pulse fidelity was not independently calibrated. The measured lifetime of the $\ket{f}$ state is approximately 400~ns, while the $\pi_{12}$ pulse duration is about 30~ns. Relaxation during the pulse can therefore limit the $\ket{e}\rightarrow\ket{f}$ transfer fidelity, contributing to the subunit slope observed in Fig.~\ref{fig:validation}(b).

\section{Assignment and bit-flip fit models}
\label{app:fit_models}

For Gaussian readout distributions with centers $\mu_g$ and $\mu_e$ and widths $\sigma_g$ and $\sigma_e$, we use
\begin{equation}
\label{eq:snr_definition}
\mathrm{SNR}
=
\frac{|\mu_g-\mu_e|}
{\sqrt{(\sigma_g^2+\sigma_e^2)/2}}.
\end{equation}
For equal variances and midpoint discrimination, the single-shot assignment error is therefore
\begin{equation}
\label{eq:assign_erfc}
p_{\mathrm{assign}}
=
\tfrac12\,\mathrm{erfc}\!\left[
\frac{\mathrm{SNR}(t_{\mathrm{RO}})}{2\sqrt{2}}
\right],\end{equation}
which equals $1-F_{\mathrm{m}}$ for the Gaussian-overlap fidelity defined in Results~\cite{blais2021circuit}.
Under white-noise integration, $\mathrm{SNR}$ grows as $\sqrt{t_{\mathrm{RO}}}$.
The RED $S_2$ channel is therefore fit in a device-dependent SNR-dominated window to
\begin{equation}
\label{eq:assign_fit}
p_{\mathrm{assign}}(t)
=
p_{\mathrm{floor}}
+
\bigl(\tfrac12-p_{\mathrm{floor}}\bigr)\,
\mathrm{erfc}\!\bigl[\beta\sqrt{\max(t-t_0,0)}\bigr],
\end{equation}
where $\beta>0$ sets the SNR growth, $t_0\ge0$ is an effective integration onset, and $0\le p_{\mathrm{floor}}<\tfrac12$ captures a residual floor.
The fit excludes $t\le t_0$ and is not extended into late-time upturns or saturation.

The branch-resolved transition channels are fit in a short-to-intermediate duration window to
\begin{equation}
\label{eq:bitflip_fit}
p_{\mathrm{bf},g}(t)=b_g+\Gamma_{\uparrow}\,t,
\qquad
p_{\mathrm{bf},e}(t)=b_e+\Gamma_{\downarrow}\,t,
\end{equation}
where $\Gamma_{\uparrow}$ describes excitation from $\ket{g}$, $\Gamma_{\downarrow}$ describes relaxation from $\ket{e}$, and $T_{\mathrm{eff}}=1/\Gamma$.
The two rates combine into an equivalent relaxation time $T_q=( \Gamma_{\uparrow}+\Gamma_{\downarrow} )^{-1}=(1/T_{\mathrm{eff}}^{\uparrow}+1/T_{\mathrm{eff}}^{\downarrow})^{-1}$, which is the quantity compared with the independently measured $T_1$ in Results.

\section{Validity and limitations of the $S_1/S_2$ decoder}
\label{app:red_recovery}

For isolated errors, the $S_1/S_2$ mapping follows directly from their effect on the repeated-readout record: a bit flip changes the physical state and produces a single broken correlation, whereas an assignment error affects only one reported outcome and therefore produces two consecutive broken correlations.

The mapping is not strictly one-to-one when multiple errors occur in neighboring cycles. For example, two adjacent bit flips produce two consecutive broken correlations and therefore mimic an $S_2$ assignment event. If bit flips occur independently with probability $p_{\mathrm{bf}}$, this contribution scales as $p_{\mathrm{bf}}^2$. Conversely, neighboring assignment errors can generate $S_1$-like events, with a contribution of order $p_a^2$. Mixed assignment and bit-flip events give additional corrections of order $p_a p_{\mathrm{bf}}$.

The syndrome probabilities can therefore be written schematically as

\begin{align}
P(S_1)
&=
p_{\mathrm{bf}}
+
O(p_a^2,p_{\mathrm{bf}}^2,p_a p_{\mathrm{bf}}),\\
P(S_2)
&=
p_a
+
O(p_a^2,p_{\mathrm{bf}}^2,p_a p_{\mathrm{bf}}).
\end{align}

Thus, the desired syndrome--channel mapping appears at first order, while contamination from overlapping errors enters at higher order. This distinction becomes important when the two error rates are strongly imbalanced. For example, when $p_a\ll p_{\mathrm{bf}}$, adjacent bit flips contribute approximately $p_{\mathrm{bf}}^2$ to the $S_2$ channel, so the absolute extraction of the assignment-error rate becomes sensitive to syndrome contamination when

\begin{equation}
p_a \lesssim p_{\mathrm{bf}}^2 .
\end{equation}

The converse occurs when the bit-flip rate is much smaller than the assignment-error rate, with contamination entering at order $p_a^2$.

\begin{figure*}[t]
  \centering
  \includegraphics[width=\textwidth]{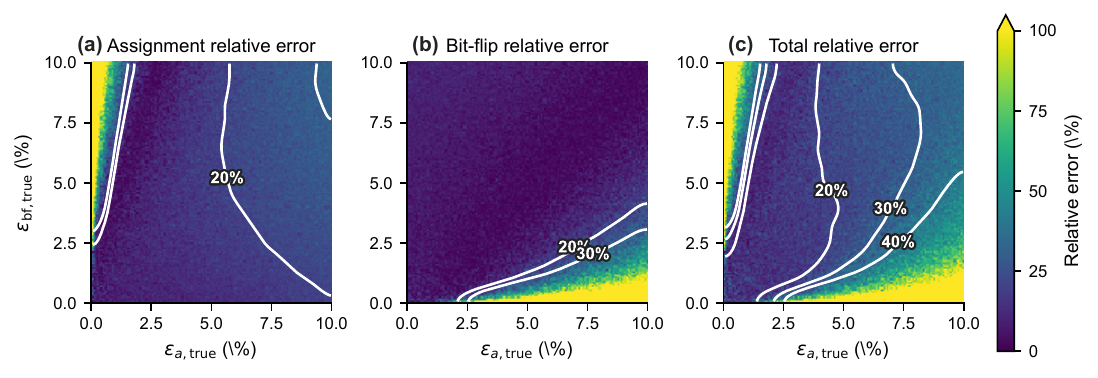}
  \caption{\textbf{Simulated RED recovery of assignment and bit-flip errors.}
  Color maps show relative recovery errors (in percent) versus true assignment $\varepsilon_{a,\mathrm{true}}$ ($x$) and true bit-flip $\varepsilon_{\mathrm{bf},\mathrm{true}}$ ($y$), also in percent, restricted to $[0,10]\%$.
  \textbf{(a)}~Assignment recovery $\varepsilon_a=\bigl|(\varepsilon_{a,\mathrm{sim}}-\varepsilon_{a,\mathrm{true}})/\varepsilon_{a,\mathrm{true}}\bigr|$.
  \textbf{(b)}~Bit-flip recovery $\varepsilon_{\mathrm{bf}}=\bigl|(\varepsilon_{\mathrm{bf},\mathrm{sim}}-\varepsilon_{\mathrm{bf},\mathrm{true}})/\varepsilon_{\mathrm{bf},\mathrm{true}}\bigr|$.
  \textbf{(c)}~Sum of the assignment and bit-flip relative recovery errors, $\varepsilon_{\mathrm{tot}}=\varepsilon_a+\varepsilon_{\mathrm{bf}}$.
  The color scale is clipped at $100\%$.}
  \label{fig:supp:red_error_maps}
\end{figure*}

We next quantify these effects using simulated repeated-readout records with input assignment- and bit-flip-error probabilities $\varepsilon_{a,\mathrm{true}}$ and $\varepsilon_{\mathrm{bf},\mathrm{true}}$. Assignment and bit-flip events are sampled independently in each readout cycle, and the resulting records are processed with the same RED decoder used for the experimental data. The extracted rates $\varepsilon_{a,\mathrm{sim}}$ and $\varepsilon_{\mathrm{bf},\mathrm{sim}}$ are then compared with the known input probabilities.

True assignment and bit-flip probabilities are swept on a $200\times200$ grid over $[0,10]\%$, with equal $\ket{g}$ and $\ket{e}$ branch rates.
Supplementary Fig.~\ref{fig:supp:red_error_maps} shows the relative recovery errors
\begin{equation}
\label{eq:supp:rel_err}
\begin{aligned}
\varepsilon_a
&=
\left|
\frac{\varepsilon_{a,\mathrm{sim}}-\varepsilon_{a,\mathrm{true}}}
     {\varepsilon_{a,\mathrm{true}}}
\right|,
\\
\varepsilon_{\mathrm{bf}}
&=
\left|
\frac{\varepsilon_{\mathrm{bf},\mathrm{sim}}-\varepsilon_{\mathrm{bf},\mathrm{true}}}
     {\varepsilon_{\mathrm{bf},\mathrm{true}}}
\right|,
\\
\varepsilon_{\mathrm{tot}}
&=\varepsilon_a+\varepsilon_{\mathrm{bf}}.
\end{aligned}
\end{equation}

Supplementary Fig.~\ref{fig:supp:red_error_maps} shows that the largest relative deviations occur when one input error rate is much smaller than the other. In this strongly imbalanced regime, higher-order syndrome mixing becomes comparable to the weaker signal, while the dominant error rate remains accurately extracted. Away from this limit, both extracted rates retain quantitative information about the underlying error channels, consistent with the higher-order mixing discussed above.

\section{Fluxonium 2 RED channels}
\label{app:fluxonium2}

\begin{figure*}[t]
  \centering
  \includegraphics[width=0.7\textwidth]{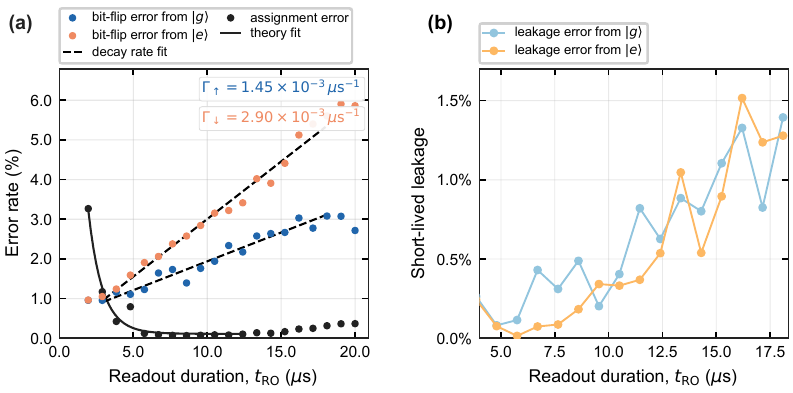}
  \caption{\textbf{Fluxonium 2 RED channels versus readout duration.}
  \textbf{(a)}~Assignment-error rate and branch-resolved bit-flip
  rates from $\ket{g}$ and $\ket{e}$. The solid black curve
  shows a Gaussian-overlap fit to the assignment error,
  while dashed lines show linear fits to the bit-flip rates,
  yielding the excitation and relaxation rates  $\Gamma_{\uparrow}$ and $\Gamma_{\downarrow}$.
  \textbf{(b)}~~Branch-resolved short-lived leakage rates
  from $\ket{g}$ and $\ket{e}$}
  \label{fig:supp:fluxonium2}
\end{figure*}

Supplementary Fig.~\ref{fig:supp:fluxonium2} shows the RED results for Fluxonium~2 as a function of readout duration.
It presents the assignment and branch-resolved bit-flip errors, together with short-lived leakage extracted from the excess $S_1$, using the same analysis as in Figs.~\ref{fig:conventional} and~\ref{fig:leakage}.

The branch-resolved bit-flip channels in Supplementary Fig.~\ref{fig:supp:fluxonium2}(a) are fit to the same linear model used in the main text [Eq.~\eqref{eq:bitflip_fit}].
The extracted rates are
$T_{\uparrow/\downarrow}=1/\Gamma_{\uparrow/\downarrow}=688/345\,\mu\mathrm{s}$,
corresponding to
$T_q=(1/T_{\uparrow}+1/T_{\downarrow})^{-1}=230\,\mu\mathrm{s}$.
This value is substantially shorter than the independently measured $T_1=1140\pm140\,\mu\mathrm{s}$ (Table~\ref{tab:device_params}).
The discrepancy is larger than for Fluxonium~1, suggesting larger readout-induced bit-flip errors on this device.
The qualitative comparison with idle $T_1$ is nevertheless the same as that reported for Fluxonium~1 and the transmon in the main text.

\section{Device parameters}
\label{app:device_params}

\begin{table}[!htb]
\caption{Device parameters and coherence time of the two fluxonium devices and one transmon device used in this work.
The data in Figs.~\ref{fig:conventional} and~\ref{fig:leakage} are from fluxonium~1 and the transmon, while the fluxonium~2 results are presented in Supplementary Fig.~\ref{fig:supp:fluxonium2}.}
\label{tab:device_params}
\begin{ruledtabular}
\begin{tabular}{lccc}
  Parameter &
  Fluxonium 1 &
  Fluxonium 2 &
  Transmon \\
  \hline
  $E_J$ (GHz) &
  8.392 &
  7.590 &
  17.2 \\
  $E_C$ (GHz) &
  1.316 &
  1.350 &
  0.169 \\
  $E_L$ (GHz) &
  0.789 &
  0.465 &
  -- \\
  $f_{01}$ &
  $104.2\,\mathrm{MHz}$ &
  $92.3\,\mathrm{MHz}$ &
  $4.65\,\mathrm{GHz}$ \\
  $f_{\mathrm{RO}}$ (GHz) &
  6.88 &
  7.21 &
  21.31 \\
  $g$ (GHz) &
  0.101 &
  0.078 &
  0.481 \\
  $T_1$ ($\mu$s) &
  $2130\pm360$ &
  $1140\pm140$ &
  $274 \pm30$ \\
  $T_{2\mathrm{E}}$ ($\mu$s) &
  $356\pm99$ &
  $309\pm103$ &
  $500 \pm60$ \\
\end{tabular}
\end{ruledtabular}
\end{table}

Table~\ref{tab:device_params} summarizes the circuit, readout, and coherence parameters of the three devices used in this work.
Results for fluxonium~1 and the transmon are presented in the main text, while the fluxonium~2 results are shown in Supplementary Fig.~\ref{fig:supp:fluxonium2}.

\clearpage

\bibliography{references_v6}

\end{document}